\documentclass[12pt,preprint]{aastex}

\newcommand{\bfb}{\mbox{\boldmath $b$}}
\newcommand{\bff}{\mbox{\boldmath $f$}}
\newcommand{\bfk}{\mbox{\boldmath $k$}}

\newcommand{\bfv}{\mbox{\boldmath $v$}}
\newcommand{\bfw}{\mbox{\boldmath $w$}}
\newcommand{\bfx}{\mbox{\boldmath $x$}}
\newcommand{\bfB}{\mbox{\boldmath $B$}}
\newcommand{\bfF}{\mbox{\boldmath $F$}}
\newcommand{\bfW}{\mbox{\boldmath $W$}}
\newcommand{\bnabla}{\mbox{\boldmath $\nabla$}}

\newcommand{\cendot}{\mbox{\boldmath $\cdot\,$}}                                        

\slugcomment{invited review, submitted to Astronomische Nachrichten}

\shorttitle{MHD Turbulence}
\shortauthors{S. Sridhar}

\begin{document}

%\twocolumn
%[

\title{Magnetohydrodynamic turbulence in a strongly magnetised plasma}  
\author{S. Sridhar}
\affil{Raman Research Institute, Sadashivanagar, Bangalore 560 080, INDIA}
\email{ssridhar@rri.res.in}

\begin{abstract}
I present a review of incompressible magnetohydrodynamic (MHD) turbulence in a strongly magnetised plasma. 
The approach is phenomenological even where a more rigorous theory is available,  so that a reader armed with paper, pencil 
and some determination may be able to work through most of the physics. The focus is on the inertial--range spectra for very large 
(fluid and magnetic) Reynolds numbers. These theories of the inertial--range are built on two important facts: (i) Kraichnan's insight that 
the turbulent cascades are a result of nonlinear interactions between oppositely directed wavepackets of Elsasser fields; (ii) these oppositely directed 
wavepackets do not exchange energy, but contribute only to changing each other's spatial structures. I begin with a description and critique of the 
Iroshnikov--Kraichnan theory, and explore the fundamental departures necessitated by the anisotropic nature of the turbulence. 
Derivations of the inertial--range spectra of four regimes of MHD turbulence --- the balanced weak, balanced strong, imbalanced weak and 
the imbalanced strong cascades --- are then presented. The need for studying the spectra of imbalanced turbulence when the waves on the 
outer scale have a short correlation time is briefly discussed.
\end{abstract}


%]

\section{Introduction}

Astrophysical systems offer numerous settings for turbulence in magnetised plasmas, such 
as the Sun, the solar wind, accretion discs, the ionised interstellar medium, molecular clouds and clusters of galaxies
(Armstrong, Rickett \& Spangler~1995, Horbury~1999, Biskamp~2003, Kulsrud~2005, Vogt \& En\ss lin~2005, 
Schekochihin \& Cowley~2007 and McKee \& Ostriker~2007). Magnetohydrodynamics (MHD)
offers the simplest description of the dynamics of a magnetised plasma. MHD is a single fluid description 
of a conducting magnetised plasma. Sources for the magnetic field, $\bfB(\bfx, t)$, could 
be external or due to currents in the plasma. Lorentz forces act on the fluid and affect the evolution of the velocity field,  
$\bfv(\bfx, t)$. If fluid motion is subsonic, incompressibility is a good approximation and the system can be 
described by the two vector fields, $\bfB$ and $\bfv$. 

Stirring a magnetised plasma on spatial scale $L$ creates velocity and magnetic field perturbations whose nonlinear evolution 
gives rise to a turbulent cascade to smaller spatial scales and is ultimately dissipated by viscosity and resistivity into heat. 
Naturally occuring processes can be somewhat different from stirring; for instance, the energy sources could be distributed in some way 
in space. In the case of the solar wind, the ultimate source of energy for MHD waves is the Sun, whereas in the interstellar medium 
the energy sources could be supernova remnants located in the spiral arms of our Galaxy. In these cases, MHD waves are generated in 
some regions of space, from which they propagate away, interact and make the plasma turbulent. Modelling this inhomogeneous 
situation is not easy. So, to simplify the physics of a difficult problem, we make the standard assumption that we are dealing with a 
statistically homogeneous and stationary situation. Thus stirring is included as a source term in the Navier--Stokes equation for the
velocity field, with appropriate statistical properties. Moreover, stirring is indispensible in numerical simulations, which complement
analytical studies. 

This review aims to be a self--contained description of MHD turbulence. However, there could be
a couple of places where the reader may have to refer to other sources, and I hope I have made this clear in the text. This review could also
seem somewhat one--sided in that much of it is based on work by me and my collaborators\footnote{A topic on which I have not worked 
is imbalanced weak MHD turbulence, and the account in \S~6 is based on my reading of Lithwick \& Goldreich~(2003).}. There are, of course, other 
points of view and the interested reader should consult the literature; these could be Iroshnikov~(1963), Kraichnan~(1965), Shebalin, Matthaeus \& 
Montgomery~(1983), Goldreich \& Sridhar~(1995, 1997), Ng \& Bhattacharjee~(1996), Cho \& Vishniac~(2000), Biskamp \& M\"uller~(2000), 
Maron \& Goldreich~(2001), Cho, Lazarian \& Vishniac~(2002), Galtier et al.~(2000, 2002),  Lithwick \& Goldreich~(2003), M\"uller, Biskamp 
\& Grappin~(2003), Galtier, Pouquet \& Mangeney~(2005), Boldyrev~(2005), M\"uller \& Grappin~(2005), Lithwick, Goldreich \& Sridhar~(2007), 
Beresnyak \& Lazarian~(2008), Chandran~(2008), Perez \& Boldyrev~(2009) and Podesta \& Bhattacharjee~(2009).

In this review the focus is on the inertial--range spectra of the magnetic and velocity fields in a strongly magnetized incompressible fluid, 
in the limit of very large (fluid and magnetic) Reynolds numbers.  \S~2 begins with the equations of incompressible MHD which govern the dynamics
of the magnetic field, $\bfB(\bfx, t)$, and the velocity field, $\bfv(\bfx, t)$. Both stirring and dissipation are included in the equations but will be dropped
later. This is because in many cases they do not directly influence the inertial--range spectra for very large Reynolds 
numbers\footnote{Dissipation does play an important role in the case of imbalanced weak MHD turbulence and this is discussed in \S~6. Stirring
can affect the spectra in the imbalanced strong cascade and this is discussed in \S~8.}. Alfv\'en and Slow waves emerge as linear perturbations
of a uniformly magnetised plasma, and their polarisations and dispersion relations are discussed. Nonlinear waves are introduced and 
their role in enabling a broad physical picture of MHD turbulence is sketched. \S~3 discusses  the Iroshnikov--Kraichnan theory of MHD turbulence and 
the problems affecting this theory. Attempts at remedying the defects lead to new physics and the next few sections are devoted to the theories that
have emerged over the last fifteen years. However, I have dispensed with the chronological in favour of a more logical presentation: thus \S~4 and 5 are on 
balanced weak and strong cascades, respectively, followed by  imbalanced weak and cascades in \S~6 and 7, respectively. In \S~8 I  briefly discuss the 
need for studying the spectra of imbalanced turbulence when the waves on the outer scale have a short correlation time.  \S~9 offers comments on matters 
that have not been discussed in the text. 

\section{The physical system and MHD turbulence}

\subsection{Incompressible MHD}

The equations describing incompressible MHD are given by,

\begin{eqnarray}
\bnabla\cdot\bfB = 0\,,\qquad \bnabla\cdot\bfv &=& 0\label{incompress}\\[2ex]
\partial_t\bfB \;-\; \bnabla\times\left(\bfv\times\bfB\right) &=& \eta \bnabla^2\bfB
\label{inductioneqn}\\[2ex]
\partial_t\bfv \;+\; \left(\bfv\cdot\bnabla\right)\bfv
\;=\; \;-\; \bnabla p
&+& \frac{\left(\bfB\cdot\bnabla\right)\bfB}{4\pi\rho}
\;+\; \nu\bnabla^2\bfv \;+\; \bff \label{nseqn}
\end{eqnarray}

\noindent
where $\rho$ is the uniform and constant mass density of the fluid, $\eta$ is the resistivity, and $\nu$ is 
the kinematic viscosity. $\bff(\bfx, t)$ is the rate of velocity stirring, assumed incompressible, $\bnabla\cendot\bff = 0$.
In many cases of interest to problems in turbulence, $\bff$ is stochastic, and its statistics is assumed to be given.
$p$ is the ratio of the total pressure to the mass density, which is determined by requiring $\bnabla\cendot\bfv = 0$. Thus we need to solve
a Poisson equation, 

\begin{equation}
\bnabla^2 p = \bnabla\cendot\left[\frac{\left(\bfB\cdot\bnabla\right)\bfB}{4\pi\rho} \;-\;  \left(\bfv\cdot\bnabla\right)\bfv\right]
\label{poisson}
\end{equation}
\noindent
with appropriate boundary conditions to determine $p$. Note that, in this incompressible limit, the thermodynamic orgins of $p$
are not so apparent. Equations~(\ref{incompress})---(\ref{poisson}), together with initial and boundary conditions, completely determine the 
evolution of the vector fields, $\bfB(\bfx, t)$ and $\bfv(\bfx, t)$. Equivalently, the system can also be described by the {\it Elsasser fields}, 
$\bfW^{\pm}$, defined by

\begin{equation}
\bfW^+ \;=\;  \bfv \;-\; \frac{\bfB}{\sqrt{4\pi\rho}}\,,\qquad
\bfW^- \;=\;  \bfv \;+\; \frac{\bfB}{\sqrt{4\pi\rho}}
\end{equation}

\noindent
When $\rho$ is uniform, we have $\bnabla\cdot\bfW^{\pm} = 0$. It turns out to be very useful to describe incompressible MHD turbulence
in terms of the Elsasser fields, instead of the magnetic and velocity fields.

Below we ignore explicit consideration of stirring and dissipative terms, because our focus is on the description of
the inertial--range at high Reynolds numbers. 

\subsection{Transverse waves}

Uniformly magnetised static solution:

\begin{equation}
\bfv_0\;=\; {\bf 0}\,,\qquad\qquad \bfB_0 \;=\; B_0\hat{z}
\label{staticsoln}
\end{equation}

\noindent
In the uniformly magnetised state  the Elsasser fields, $\bfW_0^{\pm} = \mp \,V_A\hat{z}$, where  
$V_A = B_0/\sqrt{4\pi\rho}$ is the  {\it Alfv\'en speed}. When this state is perturbed, we can write,

\begin{equation}
\bfW^+ \;=\; -V_A\hat{z} \,+\,  \bfw^+   \,,\qquad\bfW^- \;=\; V_A\hat{z} \,+\, \bfw^-\,.
\label{pert}
\end{equation}

\noindent
It must be noted that the perturbations are not necessarily small. The  perturbations obey the equations

\begin{eqnarray}
\left(\partial_t + V_A\partial_z\right)\bfw^+ + \left(\bfw^-\cdot\bnabla\right)\bfw^+ &=& - \bnabla p
\nonumber\\[2ex]
\left(\partial_t - V_A\partial_z\right)\bfw^- + \left(\bfw^+\cdot\bnabla\right)\bfw^- &=& - \bnabla p
\label{perteqns}
\end{eqnarray}

\noindent
where $p$ is determined by requiring that $\bnabla\cendot\bfw^{\pm} = 0$. The  linearised equations are

\begin{eqnarray}
\left(\partial_t + V_A\partial_z\right)\bfw^+ &=& - \bnabla p\nonumber\\[2ex]
\left(\partial_t - V_A\partial_z\right)\bfw^- &=& - \bnabla p
\label{linperteqns}
\end{eqnarray}

\noindent
If we consider solutions that vanish at infinity, $\bnabla^2p =0$ implies that  $p= {\rm constant}$ for the linearised problem, 
and we have linear waves with

\begin{eqnarray}
\bfw^+ &=& \bfF^+(x,y,z-V_At)\nonumber\\[2ex] 
\bfw^- &=& \bfF^-(x,y,z+V_At)
\label{linsolns}
\end{eqnarray}

\noindent
where $\bfF^{\pm}$ are arbitrary vector functions with $\bnabla\cendot\bfF^{\pm}=0$. It is clear that 
$\bfw^+$ is a wave that translates in the positive $z$--direction with speed $V_A$, and $\bfw^-$ is a wave that translates in 
the negative $z$--direction with speed $V_A$; wave propagation is along the magnetic field lines. We can restate this in $\bfk$--space
by assigning frequencies, $\omega^{\pm}(\bfk)$ to the $\bfw^{\pm}$ waves with dispersion relations,

\begin{equation}
\omega^+(\bfk) \;=\; V_A k_z\,,\qquad \omega^-(\bfk) \;=\; -V_A k_z
\label{disprel}
\end{equation}

\noindent
Being transverse waves, we must have $\bfk\cendot\tilde{\bfF}^{\pm}(\bfk) = 0$, where $\tilde{\bfF}^{\pm}$ is the Fourier transform
of $\bfF^{\pm}\,$. In other words, for any given $\bfk$, 
the waves have two independent degrees of polarisation, constrained to lie in the plane perpendicular to $\bfk$. The component perpendicular 
to $\hat{z}$ is called the {\it Alfv\'en wave}, and the other orthogonal component is called the  {\it Slow} (magnetosonic) {\it wave}.

\subsection{Nonlinear solutions and the general nature of MHD turbulence}

It is remarkable fact that equations~(\ref{perteqns}) admit {\it exact, nonlinear solutions}: it may be verified that, when either $\bfw^+$
or $\bfw^-$ is  equal to ${\bf 0}$ at the initial time, the nonlinear terms, $\left(\bfw^-\cdot\bnabla\right)\bfw^+$ and  
$\left(\bfw^+\cdot\bnabla\right)\bfw^-$, vanish for all time. Therefore there are exact, nonlinear solutions of the form,

\begin{equation}
\bfw^{\pm} \;=\; \bfF^{\pm}(x,y,z \mp V_At)\,,\qquad\qquad \bfw^{\mp}\;=\; {\bf 0}
\label{nonlinsolns}
\end{equation}

\noindent
In other words, wavepackets that travel in one direction (i.e. either in the positive or negative $z$--direction) do not interact nonlinearly
among themselves. Nonlinear interactions occur only between oppositely directed wave-packets. From this fact,  Kraichnan~(1965) came to
an important conclusion:

\noindent
(i) {\it Incompressible MHD turbulence in a strongly magnetised plasma can be described as being the result of (nonlinear) interactions between 
oppositely directed wavepackets}.

It is worth noting another important result that we can prove by manipulating equations~(\ref{perteqns}), namely

\begin{equation}
{d\over dt}\int{\vert \bfw^{\pm}\vert^2\over 2}\,d^3x \;=\; 0
\label{conserve}
\end{equation}

\noindent
This result is true even in the presence of nonlinearity, and is equivalent to the conservation of total energy and cross--helicity. 
Our second important conclusion now follows from equation~(\ref{conserve}):

\noindent
(ii) {\it Wavepacket collisions conserve $\pm$ve energies separately, and collisions can only redistribute the respective energies in $\bfk$--space. 
This is true regardless of how imbalanced the situation is. For instance, if stirring puts much more energy into $+$ve waves than $-$ve waves, the $+$ve
waves cannot transfer any part of their energy to $-$ve waves. They can only scatter the $-$ve waves and redistribute them (and vice--versa) in 
$\bfk$--space}. 

\section{Iroshnikov--Kraichnan (IK) Theory}

It is very useful to begin with an account of a theory of  incompressible MHD turbulence in a strongly magnetised plasma  
due to Iroshnikov~(1963) and Kraichnan~(1965), even if current views hold it to be incorrect. The physical picture is more transparent in the latter work and our 
description below follows this. Assume statistically steady, isotropic excitation of $\pm$ve waves, both with root--mean--squared (rms) amplitudes  
$w_L \ll V_A$, correlated on spatial scale  $L$, which can be referred to as the {\it stirring scale} or {\it outer scale}. Nonlinear interactions between 
oppositely directed wavepackets create $\pm$ve wavepackets on smaller spatial scales. So let us consider the nature of collisions between wavepackets 
of scale $\lambda \ll L$, where $\lambda$ is still much larger than the dissipation scale. We will assume that the most effective collisions occur between 
wavepackets of similar sizes, an assumption that is also referred to as {\it locality of interactions}. From equations~(\ref{perteqns}) it follows that, 
in one collision between a $+$ve packet and a $-$ve packet, the rms amplitude of either packet is perturbed by an amount,

\begin{equation}
\delta w_\lambda \;\sim\; {w_\lambda ^2\over V_A}
\label{ikrmsch}
\end{equation}

\noindent
We imagine that a $+$ve wavepacket of size $\lambda$ goes on to suffer a number of collisions with $-$ve wavepackets of similar scale, and vice versa.
Successive perturbations add with random phases, so that in $n$ collisions the rms amplitude of the perturbation will increase to 
$n^{1/2}(w_\lambda ^2/V_A)$. Therefore, the number of collisions for perturbations to build up to order unity is

\begin{equation}
N_\lambda \;\sim\; \left({V_A\over w_\lambda}\right)^2 
\label{iknumber}
\end{equation}

\noindent
The cascade time is the time taken for collisions to build upto order unity. Since 
each collision takes time $\sim \lambda/V_A$, the cascade time is, 

\begin{equation}
t_\lambda \;\sim\; N_\lambda\,{\lambda\over V_A}
\label{ikcastime}
\end{equation}

\noindent
We now use Kolmogorov's hypothesis that the flux of energy cascading through the inertial--range is independent of
the particular spatial scale belonging to the inertial--range. i.e. we assume that the energy flux,

\begin{equation}
\varepsilon \;\sim\; {w_\lambda^2\over t_\lambda}
\label{ikenflux}
\end{equation}

\noindent
is independent of $\lambda$. This implies that in the inertial--range,

\begin{equation}
w_\lambda \;\sim\; w_L\left({\lambda\over L}\right)^{1/4}
\label{ikscaling}
\end{equation}

\noindent
Thus we have determined the $\lambda$--dependence of the rms amplitude in the inertial--range. The same result can also be 
expressed in $\bfk$--space in terms of the (isotropic) energy spectrum, $E(k)k^3 \sim  w_\lambda^2$, where $k \sim 1/\lambda$:

\begin{equation}
E(k) \;\sim\; {w_L^2\over L^{1/2}\, k^{7/2}}
\label{ikenergy}
\end{equation}

However, we are not quite finished with our task. In our estimates, we assumed that $N_\lambda\gg 1$ and we need to verify if this is true. 
Using equations~(\ref{iknumber}) and (\ref{ikscaling}), we can estimate that  $N_\lambda \propto \lambda^{-1/2}$. Therefore $N_\lambda$ increases
as $\lambda$ decreases; in other words, $N_\lambda$ increases as we go deeper into the inertial--range. So, it looks as if our phenomenological theory 
of the IK cascade is self--consistent, and we might be tempted to conclude that the inertial--range spectrum of MHD turbulence is given by    
equation~(\ref{ikenergy}). However, the physics is not so simple. 

\subsection{Problems with the IK theory}

When  $N_\lambda\gg 1$, nonlinear interactions between oppositely directed wavepackets are weak, and we should be able to
calculate better than we have done above. There is a well--developed theory, called {\it weak turbulence}, that describes weakly nonlinear interactions 
between nearly linear waves --- see Zakharov, L'vov \& Falkovich~(1992). In this approach, perturbation theory is used to derive kinetic equations 
in $\bfk$--space describing wave--wave interactions. Then the inertial--range energy spectra emerge as stationary solutions carrying a flux of  
energy to large $k$. The lowest order of perturation theory, corresponding to what are called {\it 3--wave interactions} in weak turbulence, is 
what is needed to make rigorous the IK cascade. Here, the rate of change of energy (or wave action) at a given $\bfk$ is due to either (i) the coalescence 
of two other waves with wavevectors $\bfk_1$ and $\bfk_2$, or (i) decay of $\bfk$ into two waves with wavevectors $\bfk_1$ and $\bfk_2$. For any given 
$\bfk$, all $\bfk_1$ and $\bfk_2$ contribute, so long as they satisfy certain resonance conditions that depend on the dispersion relations satisfied by the 
linear waves. In the case of incompressible MHD, these {\it 3--wave resonance conditions} are somewhat special, because of the 
existence of the exact, nonlinear solutions given in equation~(\ref{nonlinsolns}). Since there are nonlinear interactions only between oppositely directed 
wavepackets, it means that  $\bfk_1$ and $\bfk_2$ must belong to oppositely directed wavepackets. 
Then the 3--wave resonance conditions may be written as,
 
\begin{equation}
\bfk_1 + \bfk_2 \;=\; \bfk_3\,,\qquad\omega_1^{\pm} + \omega_2^{\mp}
\;=\;\omega_k^{\pm}\quad\mbox{or}\quad \omega_k^{\mp}
\label{rescond}
\end{equation}

\noindent
where  $\omega_1^{\pm} = \pm V_A k_{1z}$,   $\omega_2^{\mp} = \mp V_A k_{2z}$,  $\omega_k^{\pm} = \pm V_A k_z$ and 
$\omega_k^{\mp} = \mp V_A k_z$. Note the resemblence to ``momentum--energy'' relations in quasi--particle interactions in 
quantum field theory or condensed matter physics. These resonance conditions have a remarkable property, noted by 
Shebalin, Matthaeus \& Montgomery~(1983), namely that one of $k_{1z}$ or $k_{2z}$ must be zero. This fact may be readily seen by considering the $z$--component 
of the wavevector resonance condition, $k_{1z} + k_{2z} = k_z$, together with the frequency resonance condition. If, for instance, we choose
$\bfk_1$ and $\bfk$ to be  $+$ waves and $\bfk_2$ to be a $-$ wave, then the frequency resonance condition is, $k_{1z} - k_{2z} = k_z$. In 
this case, it is clear that $k_{2z} = 0$, implying that  $k_{1z} = k_z$. 

\noindent
{\it Hence waves with values of $k_z$ not present initially cannot be created by nonlinear interactions during wavepacket collisions. In other words, 
there is no cascade of energy to smaller spatial scales in the $z$--direction, and hence the turbulence must be  anisotropic.} 

The somewhat surprising fact is that, attempting to create an anisotropic version of the IK theory, leads us very far from the IK theory and to entirely 
new physics. Before we get into details, it is useful to introduce some terminology. When the random stirring on the large scale $L$ generates $+$ve 
and $-$ve waves with equal rms amplitudes (or energy input), we will refer to the situation as {\it balanced}; otherwise, the turbulence is called 
{\it imbalanced}. The case of imbalanced turbulence is also refererred to as turbulence with non zero {\it cross helicity}, or non zero 
$\left<\bfv\cendot\bfb\right>$, where $\bfb$ is the magnetic field perturbation. Also, the turbulence will be called {\it weak} 
when perturbation theory is applicable (when the cascade time is larger than the collision time); otherwise the turbulence will be called {\it strong}. 
Thus there are four cases to consider: {\it balanced \& weak},  {\it balanced \& strong}, {\it imbalanced \& weak} and {\it imbalanced \& strong}. 
Weak turbulence, whether balanced or imbalanced, can be described rigorously using the kinetic equations of weak turbulence.
However strong turbulence, balanced or imbalanced, does not permit such a rigorous description and we are limited to 
phenomenological descriptions and numerical simulations.

\section{Balanced Weak MHD Turbulence}

The cascade is due to the resonant 3--wave interactions discussed above; see Ng \& Bhattacharjee~(1996), Goldreich \& Sridhar~(1997), 
Galtier et al.~(2000) and Lithwick \& Goldreich~(2003). Consider again steady, balanced and isotropic 
excitation of $\pm$ve waves, with amplitudes $w_L \ll V_A$ and scale $L$, which initiates a weak cascade to smaller spatial scales. As we have discussed 
above, there is no transfer of energy to smaller spatial scales in the direction parallel to the mean magnetic field. No parallel cascade implies that wavepackets 
of transverse scale $\lambda\ll L$ have parallel scales $L$. 

It turns out that the different polarisations of the Alfv\'en waves and Slow waves play a very important role in the turbulent cascade.
Viewed physically, we may think of the $\left(\bfw^{\pm}\cdot\bnabla\right)\bfw^{\mp}$  terms as being responsible for the nonlinear 
cascade\footnote{This statement is not entirely accurate, because the $\bnabla p$ term also contributes a quantity of the same order of magnitude 
which does not have this precise form. However, the argument advanced in the text suffices for our phenomenological account in this review.}.
During collisions between two oppositely directed wavepackets, we may think of  $\left(\bfw^{\pm}\cdot\bnabla\right)$ as coming from the wavepacket
that offers a perturbation, and the $\bfw^{\mp}$ wavepacket as the one that suffers the perturbation. For $\lambda\ll L$, the polarisation of an 
Alfv\'en wave is such that its $\bfw^{\pm}$ is  very nearly perpendicular to $\hat{z}$, whereas $\bfw^{\pm}$ in a Slow wave 
is almost parallel to $\hat{z}$. Then we estimate that
$\left(\bfw^{\pm}\cdot\bnabla\right) \sim (w_\lambda/\lambda)$ when the perturbing wavepacket is an Alfv\'en wave, and 
$\left(\bfw^{\pm}\cdot\bnabla\right) \sim (w_\lambda/L)$ when the perturbing wavepacket is a Slow wave. {\it Clearly an Alfv\'en wave is a
much stronger perturber than a  Slow wave.} Thus, it comes as no surprise that Slow waves are mostly scattered by Alfv\'en waves, but do not 
significantly perturb Alfv\'en waves; {\it  the cascade of Slow waves is slaved to the Alfv\'en waves.} Hence in our estimates below we ignore 
Slow waves, and restrict attention to Alfv\'en waves.

A collision between a $+$ve Alfv\'en wavepacket and a $-$ve Alfv\'en wavepacket now takes time $\sim L/V_A$, because the wavepackets have
parallel scales $\sim L$. In one collision a wavepacket is perturbed by amount, 

\begin{equation}
\delta w_\lambda \;\sim\; {L w_\lambda ^2\over \lambda V_A} \;<\; w_\lambda
\label{bwrmsch}
\end{equation}

\noindent
The perturbations add with random phases. The number of collisions for perturbations to build up to order unity is

\begin{equation}
N_\lambda \;\sim\; \left({\lambda V_A\over L w_\lambda}\right)^2
\label{bwnumber}
\end{equation}

\noindent
The cascade time is

\begin{equation}
t_\lambda \;\sim\; N_\lambda\,{L\over V_A}
\label{bwcastime}
\end{equation}

\noindent
Kolmogorov's hypothesis of the $\lambda$-independence of the  energy flux

\begin{equation}
\varepsilon \;\sim\; {w_\lambda^2\over t_\lambda}
\label{bwenflux}
\end{equation}

\noindent
implies that in the inertial-range, the rms amplitudes are given by,

\begin{equation}
w_\lambda \;\sim\; w_L\left({\lambda\over L}\right)^{1/2}
\label{bwscaling}
\end{equation}

\noindent
The energy spectrum, $E(\bfk)$, is highly anisotropic in $\bfk$--space, and now depends on both $k_z$ and $k_\bot$. The precise dependence of $E$ on $k_z$ 
is not so interesting, because there is no parallel cascade; it suffices to know that $E$ is largely confined to the region $\vert k_z\vert < L^{-1}$. However, the 
dependence of $E$ on $k_\bot$ is of great interest, because this has been established by the cascade in the transverse directions, described above. 
Accounting for this anisotropy, we can estimate, $E(k_\bot, k_z)\,k_\bot^2\,L^{-1}\;\sim\; w_\lambda^2$, where $k_\bot\sim \lambda^{-1}$. Now, using 
equation~(\ref{bwscaling}), we have

\begin{equation}
E(k_\bot, k_z) \;\sim\; {w_L^2\over k_\bot^3}\,,\quad\mbox{for}\quad\vert k_z\vert < L^{-1}
\label{bwenergy}
\end{equation}

\noindent
Being a theory based on weak turbulence, the turbulent cascade can be described rigorously using kinetic equations for energy transfer;
see Galtier et al.~(2000) and Lithwick \& Goldreich~(2003).

Having obtained the inertial--range energy spectrum of the balanced weak cascade, we need to perform checks of self consistency
regarding the assumed weakness of the cascade. Using equations ~(\ref{bwnumber}) and (\ref{bwscaling}), we estimate that

\begin{equation}
N_\lambda \;\sim\; \left({V_A\over w_L}\right)^2\,{\lambda\over L}
\label{bwnumfin}
\end{equation}

\noindent
We must first verify that the cascade initiated at the stirring scale $L$ is weak to begin with, which implies that we must have $N_L\gg 1$. 
From equation~(\ref{bwnumfin}),   we see that this can be satisfied if $w_L \ll V_A$. In other words isotropic, balanced stirring initiates a
weak cascade if the rms amplitudes at the stirring scale is much less than the  Alfv\'en speed. A more geometric way of stating this is that 
a weak cascade is initiated if stirring bends field lines (of the mean magnetic field) only by small angles.  

\noindent
Equation~(\ref{bwnumfin}) implies more interesting consequences: {\it the transverse cascade  strengthens as $\lambda$ decreases}, 
because $N_\lambda$ decreases when  $\lambda$ decreases. Therefore the assumption of the weakness of interactions must break down when 
$N_\lambda \sim 1$, which happens when $\lambda\sim\lambda_*\sim L(w_L/V_A)^2$. Therefore, the inertial--range spectrum of 
equation~(\ref{bwenergy}) is valid only for transverse scales larger than $\lambda_*$.

\section{Balanced Strong MHD Turbulence}

When $N_\lambda \sim 1$, the assumption of the weakness of the cascade is no longer even approximately valid. The turbulent cascade 
turns strong, and was first described by Goldreich \& Sridhar~(1995). They conjectured that $N_\lambda$ remains of order unity for smaller values of 
$\lambda$ all the way down to the dissipation scale. This is equivalent to assuming that the cascade time remains of order the wave period, 
a condition that may be referred to as {\it critical balance}. The balanced weak cascade described above turns into a balanced strong cascade 
for $\lambda < \lambda_*$. Instead of following the transition from weak to strong, we prefer to describe the inertial--range of the balanced strong 
cascade when it is initiated at the stirring scale itself because there is less clutter in the description. 

Consider steady, balanced and isotropic excitation of $\pm$ve waves, with amplitudes $w_L \sim V_A$ and scale $L$, which initiates a strong cascade 
to smaller spatial scales. Wavepackets of transverse scale $\lambda < L$ can now possess parallel scales that are smaller than $L$. This is because
the resonance conditions are not in force when nonlinear interactions are strong so, in addition to a transverse cascade, a parallel cascade can also occur.
Let eddies of transverse scale $\lambda$ possess  parallel scale $\Lambda_\lambda$, which is as yet an unknown function 
of the transvarse scale $\lambda$. Critical balance implies that

\begin{equation}
t_\lambda \;\sim\; {\lambda\over w_\lambda}\;\sim\; {\Lambda_\lambda\over V_A}
\label{bscrit}
\end{equation}

\noindent
As before we invoke Kolmogorov's hypothesis of the $\lambda$--independence of the energy flux

\begin{equation}
\varepsilon \;\sim\; {w_\lambda^2\over t_\lambda}
\label{bsenflux}
\end{equation}

\noindent
which implies that in the inertial--range

\begin{equation}
w_\lambda \;\sim\; V_A\left({\lambda\over L}\right)^{1/3}\,,\qquad
\Lambda_\lambda\;\sim\; L^{1/3}\lambda^{2/3}
\label{bsscaling}
\end{equation}

\noindent
The energy spectrum is again anisotropic and may be estimated as,
$E(k_\bot, k_z)\,k_\bot^2\,\left(\Lambda_\lambda\right)^{-1}\;\sim\; w_\lambda^2$, where $k_\bot\sim \lambda^{-1}$. Now, using 
equation~(\ref{bsscaling}), we have

\begin{equation}
E(k_\bot, k_z) \;\sim\; {w_L^2\over L^{1/3}k_\bot^{10/3}}\,,\quad\mbox{for}\quad\vert k_z\vert < {k_\bot^{2/3}\over L^{1/3}}
\label{bwsenergy}
\end{equation}

\noindent
{\bf Summary of the balanced strong cascade}

\noindent
1. The cascade is {\it critically balanced} in that the cascade time is of order of the wave period throughout the inertial--range.

\noindent
2. The rms amplitudes have a scaling with the transverse scale which is of the Kolmogorov form. 

\noindent
3. The turbulent cascade occurs in the transverse as well as parallel directions. However, the cascade occurs predominantly in the transverse 
directions, because the parallel correlation lengths scale as the $2/3$rd power of the transverse correlation lengths.

\section{Imbalanced Weak MHD Turbulence}

The cascade is due to the resonant 3--wave interactions, see  Galtier et al.~(2000) and Lithwick \& Goldreich~(2003).
Consider steady, imbalanced and isotropic 
excitation of $\pm$ve Alfv\'en waves, with rms amplitudes $w^-_L \leq w^+_L \ll V_A$ and scale $L$, which initiates weak cascades to smaller spatial scales. 
As we have discussed earlier, the 3--wave resonance conditions forbid a parallel cascade, so wavepackets of transverse scale $\lambda\ll L$ have 
parallel scales $L$. 

\noindent
The cascade times for the $\pm$ve wavepackets can be estimated by considerations similar to those for the 
balanced case. Consider collisions suffered by a $+$ve wavepacket of transverse scale $\lambda$ and parallel scale $L$, with 
$-$ve wavepackets of similar spatial scales. A single collision occurs over time $\sim L/V_A$, during which 
the $+$ve wavepacket is perturbed by amount, $\delta w_\lambda^+ \sim (L w_\lambda^- w_\lambda^+/ \lambda V_A)$.
Successive perturbations add with random phases, so the number of collisions for perturbations to build up to order unity is
$N_\lambda^+ \sim (\lambda V_A/ L w_\lambda^-)^2$. Similar considerations apply to collisions suffered by a $-$ve wavepacket  
with many $+$ve wavepackets, and hence $N_\lambda^- \sim (\lambda V_A/ L w_\lambda^+)^2$. Therefore the 
cascade times for the $\pm$ve waves are, 

\begin{equation}
t_\lambda^+ \;\sim\; \frac{V_A}{L}\left(\frac{\lambda}{w_\lambda^-}\right)^2\,,\qquad
t_\lambda^- \;\sim\; \frac{V_A}{L}\left(\frac{\lambda}{w_\lambda^+}\right)^2
\label{ibwcastime}
\end{equation}

\noindent
Kolmogorov's hypothesis of the $\lambda$--independence of the  energy fluxes

\begin{equation}
\varepsilon^+ \;\sim\; {\left(w_\lambda^+\right)^2\over t_\lambda^+}\,,\qquad
\varepsilon^- \;\sim\; {\left(w_\lambda^-\right)^2\over t_\lambda^-}
\label{ibwenflux}
\end{equation}

\noindent
imples that

\begin{equation}
w_\lambda^+w_\lambda^-\propto \lambda
\label{ibwrel}
\end{equation}

\noindent
In other words, the requirement that both $\varepsilon^+$ and $\varepsilon^-$ be independent of $\lambda$ leads to just one  
relation for the two quantities, $w_\lambda^+$ and $w_\lambda^-$. 

\noindent
{\it Hence scaling arguments are, by themselves, insufficient for the determination of the inertial--range spectra of the imbalanced cascade.  
The new physics that determines the spectra is the fact that the $\pm$ energies are forced to equalise at the dissipation scale; 
this  pinning completely determines the spectra}. 

\noindent
Let the dissipation scale be at $\lambda_{\rm dis}$, and the common rms amplitude of $w_\lambda^+$ and $w_\lambda^-$ at this 
scale be $w_{\rm dis}$. Then equation~(\ref{ibwrel}) implies that we can write,

\begin{eqnarray}
w_\lambda^+ &=& w_{\rm dis}\left(\frac{\lambda}{\lambda_{\rm dis}}\right)^{(1+\alpha)/2}\nonumber\\[2ex]
w_\lambda^- &=& w_{\rm dis}\left(\frac{\lambda}{\lambda_{\rm dis}}\right)^{(1-\alpha)/2}
\label{ibwcaling}
\end{eqnarray}

\noindent
where the parameters,  $w_{\rm dis}$, $\lambda_{\rm dis}$ and $\alpha$ are to be determined. When
the dissipation is caused by diffusive processes and $\nu\sim \eta$ (i.e. the Prandtl number is close to unity), 
the dissipation timescale is

\begin{equation}
t_{\rm dis} \;\sim\; \frac{\lambda_{\rm dis}^2}{\nu}
\label{ibwdistime}
\end{equation}
 
\noindent
At the dissipation scale, the cascade time and the dissipation time are comparable. Setting $\lambda\sim\lambda_{\rm dis}$ in 
equation~(\ref{ibwcastime})  and setting it equal to $t_{\rm dis}$, we can estimate that,

\begin{equation}
w_{\rm dis} \;\sim\; \left(\frac{\nu V_A}{L}\right)^{1/2}
\label{ibwwdis}
\end{equation}

\noindent
Note that $w_{\rm dis}$ is independent of the the energy fluxes, $\varepsilon^{\pm}$. However, the other two parameters, 
$\lambda_{\rm dis}$ and $\alpha$ depend on the energy fluxes, and these can be determined only by use of the kinetic equations
of the weak turbulence theory of the cascade, as shown by  Galtier et al.~(2000) and Lithwick \& Goldreich~(2003)\footnote{It should be noted that $\lambda_{\rm dis}$ 
and $\alpha$ can be expressed in terms of the rms amplitudes at the stirring scale, $w_L^+$ and $w_L^-$, but it is the energy fluxes, 
$\varepsilon^{\pm}$, that are the more interesting physical quantities.}. {\it So, in contrast to the balanced weak cascade, dissipation plays 
a direct role in determing the inertial--range spectra of imbalanced weak MHD turbulence}.

In common with the balanced weak case, we still need to verify that the turbulence is indeed weak in the inertial--range, 
So we need to verify that the quantitites, 

\begin{equation}
\chi^\pm_\lambda \;\equiv\; {L w^\pm_\lambda\over \lambda V_A} \;\ll\; 1
\label{ibwchi}
\end{equation}

\noindent
throughout the inertial--range, $L > \lambda > \lambda_*$. To do this, we use $w_\lambda^+w_\lambda^-/\lambda = w_{\rm dis}^2/\lambda_{\rm dis}$
and equation~(\ref{ibwwdis}) to first work out $\lambda_{\rm dis}$: 

\begin{equation}
\lambda_{\rm dis} \sim \frac{\nu V_A}{w_L^+w_L^-}
\label{ibwlambdadis}
\end{equation}

\noindent
For the inertial--range to be of large enough extent we must have $L \gg \lambda_{\rm dis}$, which implies that

\begin{equation}
\frac{w_L^+w_L^-}{V_A^2} \;\gg\; \frac{\nu}{LV_A}
\label{ibwcond1}
\end{equation}

\noindent
Now, requiring that $\chi^\pm_\lambda \ll 1$ when $ \lambda \sim \lambda_*$ gives us another condition:

\begin{equation}
\frac{w_L^+w_L^-}{V_A^2} \;\ll\; \left(\frac{\nu}{LV_A}\right)^{1/2}
\label{ibwcond2}
\end{equation}

\noindent
When $\nu\ll LV_A$, the above inequalities can be satisfied with appropriately chosen $w_L^+w_L^-$, and an imbalanced weak cascade
with a substantial inertial--range is possible.

\section{Imbalanced Strong MHD Turbulence}

This is clearly the more general and generic case of MHD turbulence, and the account below is based on Lithwick, Goldreich \& Sridhar~(2007). 
Consider steady, imbalanced and isotropic excitation of $\pm$ve waves, with rms amplitudes $w^-_L \leq w^+_L < V_A$ and scale $L$. 

At some transverse scale $\lambda \ll L$, let the amplitudes be $w^+_\lambda \geq w^-_\lambda$, and parallel correlation lengths 
be $\Lambda^+_\lambda$ and $\Lambda^-_\lambda$. Note that we allow the parallel correlation lengths to be functions of the transverse scale, 
because resonance conditions are not in force in strong turbulence. $+$ve waves bend field lines by angle $(w^+_\lambda/V_A)$, hence a 
$+$ve wavepacket of parallel scale $\Lambda^+_\lambda$ will displace field lines in the transverse directions by distances 
$(w^+_\lambda\Lambda^+_\lambda/V_A)$. If this distance exceeds the transverse correlation length, $\lambda$, then the 
 $-$ve waves will suffer strong perturbations. In other words, the $-$ve cascade is strong if

\begin{equation}
\chi^+_\lambda \;\equiv\; {\Lambda^+_\lambda w^+_\lambda\over \lambda V_A} \geq 1
\label{ibschiplus}
\end{equation}

\noindent
Then the  $-$ve wave cascade time is

\begin{equation}
t^-_\lambda \;\sim\; {\lambda\over w^+_\lambda}
\label{ibsnegcastime}
\end{equation}

\noindent
and the $-$ve wave parallel correlation length is

\begin{equation}
\Lambda^-_\lambda \;\sim\;V_A t^-_\lambda \;\sim\; \lambda\left({V_A\over w^+_\lambda}\right)
\label{ibscorrlenght}
\end{equation}

\noindent
From equations~(\ref{ibschiplus}) and (\ref{ibscorrlenght}), we note that 

\begin{equation}
\Lambda^+_\lambda \;\geq\; \Lambda^-_\lambda
\label{ibsunequal}
\end{equation}

\noindent
We can infer a deeper result. Any two regions in the $+$ve packet that are separated by parallel lengths $>\Lambda^-_\lambda$ will be 
cascaded by statistically uncorrelated $-$ve waves. {\it Hence the $-$ve waves will imprint their parallel scales on the $+$ve waves, and 
both $\pm$ve waves will have the same parallel correlation lengths, $ \Lambda_\lambda\,$:}

\begin{equation}
\Lambda^+_\lambda \;\sim\; \Lambda^-_\lambda \;\sim\; \Lambda_\lambda  \;\sim\; \lambda\left({V_A\over w^+_\lambda}\right)
\label{ibsequal}
\end{equation}

\noindent
From equations~(\ref{ibschiplus}) and (\ref{ibsequal}) we have, 

\begin{equation}
\chi^+_\lambda \;\sim\; 1
\label{ibsnegcrit}
\end{equation}

\noindent
{\it Therefore the $-$ve cascade is  critically balanced}.

It turns out that the  $+$ve wave cascade time is,

\begin{equation}
t^+_\lambda \;\sim\; {\lambda\over w^-_\lambda}
\label{ibsposcastime}
\end{equation}

\noindent
This means that the straining rate imposed by the $-$ve waves on the $+$ve waves,  $(w^-_\lambda/\lambda)$, is imposed 
coherently over time $(\lambda/ w^-_\lambda)$. Then

\begin{equation}
{\mbox{ $-$ve wave coherence time}\over\mbox{$-$ve wave wave period}} \sim 
{t^+_\lambda V_A\over\Lambda_\lambda} \sim {w^+_\lambda \over w^-_\lambda} \geq 1
\label{ibsratio}
\end{equation}

\noindent
How can the coherence time of the $-$ve waves exceed their wave period? The answer is involved and is worked out in 
the Appendix of Lithwick, Goldreich \& Sridhar~(2007); here we merely summarise the result. Now let us look at the interaction between $\pm$ve waves of 
transverse scale $\lambda$ and parallel scale $\Lambda_\lambda$, {\it in the rest frame of the $+$ve waves}. In this frame, the $-$ve waves 
alter $+$ve waves which react back onto the $-$ve waves. Let $z'$ measure distance along the parallel direction in the rest frame of 
the $+$ve waves. At a fixed $z'$ location the $+$ve waves are changing on their cascade time scale $t^+_\lambda$. Hence, 
over times separated by $t^+_\lambda$, the $-$ve waves crossing $z'$ are cascaded by entirely different $+$ve waves. This 
implies that $t^-_{{\rm corr}\lambda} \;\sim\;t^+_\lambda$. Because the $+$ve waves are strained at rate $(w^-_\lambda/\lambda)$, 
it follows that $t^+_\lambda \sim (\lambda/ w^-_\lambda)$.

We are now in a position to work out the scalings for the imbalanced strong cascade. Kolmogorov's hypothesis of the 
$\lambda$--independence of the energy fluxes

\begin{eqnarray}
\varepsilon^-  &\sim & {(w^-_\lambda)^2\over t^-_\lambda} \;\sim\; {(w^-_\lambda)^2 w^+_\lambda\over \lambda}
\nonumber\\[2ex]
\varepsilon^+ &\sim & {(w^+_\lambda)^2\over t^+_\lambda} \;\sim\; {(w^+_\lambda)^2 w^-_\lambda\over \lambda}
\label{ibsenergy}
\end{eqnarray}

\noindent
implies that in the inertial range,

\begin{eqnarray}
w^\pm_\lambda &\sim & {(\varepsilon^\pm)^{2/3}\over (\varepsilon^\mp)^{1/3}}\,\lambda^{1/3}
\nonumber\\[2ex]
\Lambda_\lambda &\sim & {(\varepsilon^-)^{1/3}\over (\varepsilon^+)^{2/3}}\,V_A\lambda^{2/3}
\label{ibsscalings}
\end{eqnarray}

\noindent
Equations~(\ref{ibsenergy}) are given in Verma et al.~(1996) and Verma~(2004), although isotropy is assumed. 

\noindent
{\bf Summary of the imbalanced strong cascade}

\noindent
1. The ratio of the Elsasser amplitudes is independent of scale, and is equal
to the ratio of the corresponding energy fluxes. Thus we can infer turbulent flux ratios 
from the amplitude ratios, thus providing insight into the origin of the turbulence.

\noindent
2. In common with the balanced strong cascade, the energy spectra of both Elsasser
waves are of the anisotropic Kolmogorov form, with their parallel
correlation lengths equal to each other on all scales, and
proportional to the 2/3rd power of the transverse correlation length. 

\noindent
3. The equality of cascade time and waveperiod (critical balance) that
characterizes the strong balanced cascade does not apply to the Elsasser field
with the larger amplitude. Instead, the more general criterion that always applies to both Elsasser fields 
is that the cascade time is equal to the correlation time of the straining imposed by oppositely-directed waves.

\noindent
4. In the limit that the energy fluxes are equal, the turbulence corresponds to the
balanced strong cascade.

We note that there are other theories of imbalanced cascades which differ from the above account; to the best of my knowledge, 
these are given in Beresnyak \& Lazarian~(2008), Chandran~(2008), Perez \& Boldyrev~(2009) and Podesta \& Bhattacharjee~(2009).

\section{Stirring--affected scales in the imbalanced strong cascade}

Consider again steady, imbalanced and isotropic excitation of $\pm$ve waves, with rms amplitudes $w^-_L \leq w^+_L < V_A$, on scale $L$.
The wave periods of the $\pm$ve waves at the stirring scale  $L/V_A$. Let the {\it coherence time} of the waves on the stirring scale be $T$; 
we expect that $T\geq L/V_A$, but otherwise $T$ is unconstrained and determined only by the physics of however the $\pm$ve waves are 
generated on scales $\sim L$. The theory of Lithwick, Goldreich \& Sridhar~(2007) applies when $T > t^+_L$; otherwise, when $T < t^+_L$, the coherent 
straining of $+$ve waves by the $-$ve waves will be interrupted before the $+$ve waves can cascade. Using,

\begin{equation}
t^+_\lambda \;\sim\;  t^-_{{\rm corr}\lambda} \;\sim\; {(\varepsilon^+)^{1/3}\over (\varepsilon^-)^{2/3}}\,\lambda^{2/3}
\label{ibsplustime}
\end{equation}

\noindent
we can infer that $T < t^+_L$ implies, 

\begin{equation}
\ell \;\equiv\; {(\varepsilon^-)\,T^{3/2} \over (\varepsilon^+)^{1/2}} \;<\; L
\label{ibselldef}
\end{equation}

\noindent
Then, the results of Lithwick, Goldreich \& Sridhar~(2007) are valid for only for transverse scales $\lambda < \ell$. 

The range of transverse scales, $\ell < \lambda < L$, can be referred to as {\it stirring--affected scales}. For these scales, 
the $+$ve wave cascade is no longer strong and $\pm$ve spectra are, as yet, unknown. It would also be interesting to see if there is any signature of these 
stirring--affected scales in solar wind turbulence.

\section{Some comments}

We have assumed that the strong large--scale magnetic field is uniform in magnitude and direction, but this need not be 
the case, as Kraichnan~(1965) realised. We could consider a more general setting in which the magnetic field is disordered on 
large scales, with the largest spatial correlation length, $L$, being smaller than the system size. Let the magnetic energy
in eddies of size $L$ be larger than the magnetic energy on smaller scales. Then the magnetic field of the eddies of size $L$
will act as the mean magnetic field for Alfv\'en and Slow waves of smaller scales, and all of what we have derived in this review 
for the inertial--range of  MHD turbulence in a strong magnetic field may be expected to be still valid.\footnote{There is an interesting 
detail here that is, perhaps, worth noting. In the case of hydrodynamic turbulence, the velocity field of the large--scale eddies (of size $L$) 
will ``sweep'' eddies of smaller sizes, and this effect gives rise to observational consequences for the velocity time correlation function measured 
by a probe fixed at some location in the fluid. However, sweeping cannot affect the dynamics of nonlinear interactions, because a large--scale 
velocity field can be effectively transformed  away 
by an appropriate Galilean boost. On the other hand, a large--scale magnetic field cannot be similarly transformed away; one way to physically see this is 
to note that such a field enables (Alfv\'en and Slow) waves that travel both up and down the field lines, and these cannot be wished away by 
any Galilean boost.} The opposite case of the large--scale magnetic field being weak often occurs in 
problems concerning dynamo processes, and turns out to be harder to analyse.

We have considered incompressible MHD turbulence, but the general case in astrophysics involves compressive fluids. 
In compressible MHD, there are four types of waves: Alfv\'en, Fast, Slow and Entropy. The inertial--range spectra we have derived are applicable 
to cascades of Alfv\'en waves, which are always incompressible. These are valid, so long as the other three waves 
(which are compressive) have negligible effect on the Alfv\'en wave. That this is indeed so has been discussed in the following papers:
Lithwick \& Goldreich~(2001), Maron  \& Goldreich~(2001) and Cho \& Lazarian~(2003).

In the context of Navier--Stokes turbulence of a neutral fluid, it is well--known that a triple correlation function
of velocities is related to the rate of energy input; see Karman \& Howarth~(1938), Kolmogorov~(1941) and Monin~(1959). 
Similar relations in the context of MHD turbulence have been obtained by Politano \& Pouquet~(1998a,b) and Podesta~(2008). However, the relation to 
the theories of the inertial--range spectra
of MHD turbulence has not received as much attention as in the case of neutral fluids. 

It has sometimes been argued that there is a ``dynamical alignment'' of velocity and magnetic fields, resulting in spectra that are flatter than Kolmogorov ---
see Boldyrev~(2005) and Beresnyak \& Lazarian~(2006). However, observations of solar wind turbulence show spectra that are consistent with Kolmogorov 
-- see Horbury, Forman \& Oughton~(2005).

\acknowledgements
I would like to thank (i) the organisers of the meeting on Astrophysical Magnetohydrodynamics for their hospitality, (ii) 
Peter Goldreich and Yoram Lithwick for all the things I learnt from them.

\end{document}